# Parallel Self-Sorting System for Objects

Samuel King Opoku, *Member, IEEE*

*Abstract*—Conventional sorting algorithms make use of such data structures as array, file and list which define access methods of the items to be sorted. These traditional methods – exchange sort, divide and conquer sort, selection sort and insertion sort – require supervisory control program. The supervisory control program has access to the items and is responsible for arranging them in the proper order. This paper presents a different sorting algorithm that does not require supervisory control program. The objects sort themselves and they are able to terminate when sorting is completed. The algorithm also employs parallel processing mechanisms to increase its efficiency and effectiveness. The paper makes a review of the traditional sorting methods, identifying their pros and cons and proposes a different design based on conceptual combination of these algorithms. Algorithms designed were implemented and tested in Java desktop application.

## I. INTRODUCTION

Conventional sorting algorithms make use of a data structure to store the items to be sorted. The general data structures employed are array, file, list or some kind of collection which defines access methods of the items. The popular traditional sorting algorithms are grouped as insertion; divide and conquer; bubble and selection [1], [2], [14], [15]. Insertion sort works by taking elements from a list one by one and inserting them in their correct position into a new sorted list [19]. It is relatively efficient for small list and mostly sorted list [3], [4], [16]. Divide and conquer algorithms rely on partition operation. The commonest divide and conquer algorithm is the Quick-sort. Quick-sort selects a pivot element and uses it to partition the dataset. Elements that are smaller than the pivot are moved to positions before it and all elements greater than the pivot element are moved to positions after it. The most complex issue in quick-sort is choosing a good pivot element. Consistent poor choices of pivots can result in drastically slower $O(n^2)$ performance. If at each step, the median is chosen as the pivot then the algorithm works in $O(n \log n)$ [1], [3], [5]. Finding the median however, is an $O(n)$ operation on unsorted list and therefore exerts its own penalty with the sorting.

Bubble sort is a simple algorithm. The algorithm starts at the beginning of a dataset. It compares the first two elements and if the first is greater than the second, it swaps them. It continues for each pair of adjacent elements to the end of the dataset. It then starts again repeating the process until no swap occurs at the last pass. Average case and worst case are both $O(n^2)$ performance. This algorithm is efficiently used on a list that is already sorted except for a very small number of elements [1], [6], [7]. For instance, if only one element is not in order, bubble sort takes 2n time. For two elements not in order, it takes 3n time.

Selection sort algorithms are noted for its simplicity [18]. It finds the minimum value, swaps it with the value in the first position and repeats these steps for the remainder of the list. It does no more than n swaps and thus it is useful where swapping is very expensive. It has $O(n^2)$ complexity making it inefficient on large list [1], [8]. Heap sort [3], [8] is a much efficient version of selection sort. It also works by determining the largest (or smallest) element of the list placing that at the end (or beginning) of the list. It then continues with the rest of the list. The task is efficiently accomplished when a heap data structure, a special type of binary tree, is used. Using the heap data structure, finding the next largest element takes $O(\log n)$ time instead of $O(n)$ for a linear scan allowing the heap sort to run in $O(n \log n)$ time [3], [16].

Self sorting algorithm was designed in [9]. The algorithm was based on bubble sort and quick sort algorithms. The self sorting algorithm requires that the object knows all the state or status of the preceding objects and the next objects before it can determine whether all the objects are sorted or not. It therefore required finding the first object which is based on bubble sort algorithm and finding the last object which is based on quick sort algorithm [1]. The first object is the object whose reference to the previous object is null whereas the last object is an object whose reference to the last object is null. This paper presents self sorting algorithm that does not require supervisory control program. It also employs parallel processing mechanism to increase the efficiency and effectiveness of the algorithm. The algorithm is then implemented and tested using Java desktop application.

Self-sorting is the ability of objects to find and self-assemble selectively with their corresponding recognition units [10], [18]. Self sorting plays important roles in our daily lives from complex systems, such as DNA replication and transcription, to simple phenomena, such as oil-water phase separation. A self sorting object is implemented such that it has a sort method that allows the object to place itself in the correct position with respect to preceding and following objects. The sorting algorithm usually employed by self-sorting objects may root from single or combination of the known conventional sorting algorithms [9], [20]. A self sorting object has at least three features [9], [10]:

This work was supported in part by the Department of Software Engineering, ICT System Solution Center.
The author is with the Computer Science Department, Kumasi Polytechnic, Postal Code 854, Kumasi, Ghana, West Africa (phone: +233-242-124-291; e-mail: Samuelk.opoku@kpoly.edu.gh).

- Stores a reference to the preceding object or null if there is none.
- Stores a reference to the next object or null if there is none
- Stores a data value to be used as the sort key

A Java desktop application is implemented in Java 2 Standard Edition (J2SE). J2SE describes the Java language and the basic set of API libraries that are used to create window (or frame and panel as in Java) and applet applications [11]. Applets are applications that run within web browsers. A typical architecture of J2SE is illustrated in the figure below:

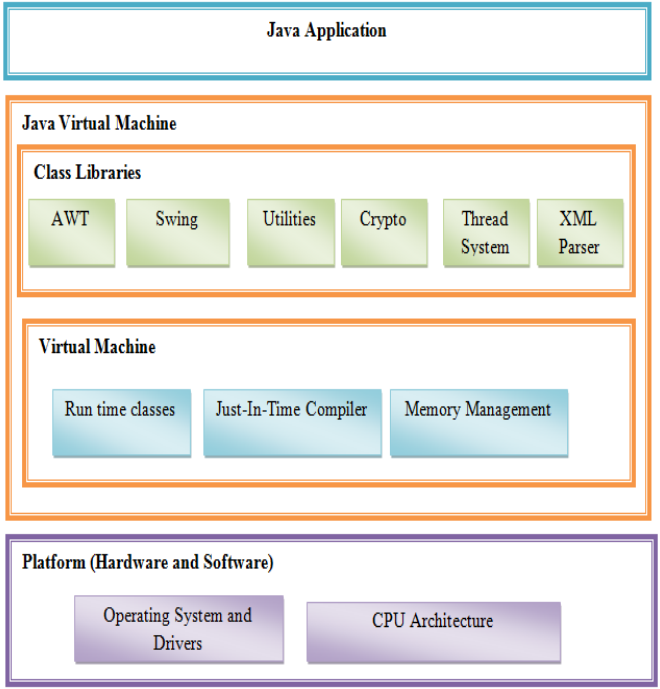

Fig. 1. J2SE Architecture

Sorting algorithms implemented in J2SE generally require that objects implement the Comparable interface [12]. The Comparable interface has just one method; int compareTo(Object obj). The implementation of the method returns int less than zero if the object comes before the given object in some natural order. Zero if the object equals to the given object and an int more than zero if the object comes after the given object [11], [12]

A parallel processing refers to the concept of speeding-up the execution of a program by dividing the program into multiple fragments that can execute simultaneously. There are two forms of parallelism [13]. These are transparent or implicit parallelism and explicit multi-process parallelism. Transparent parallelism breaks a job into parallel threads without the intervention of the user whereas explicit multi-process parallelism requires users to formulate and break job in terms of both function and data. Parallel processing has the following characteristics [13], [17], [20]:

- It includes job scheduling and other serial computation
- A basic loop starts with supervisory scheduling followed by the computation and inter-threading message phases
- Synchronization occurs prior to returning to scheduling the next unit of parallel work.

The figure below illustrates the general overview of the architecture of parallel processing system.

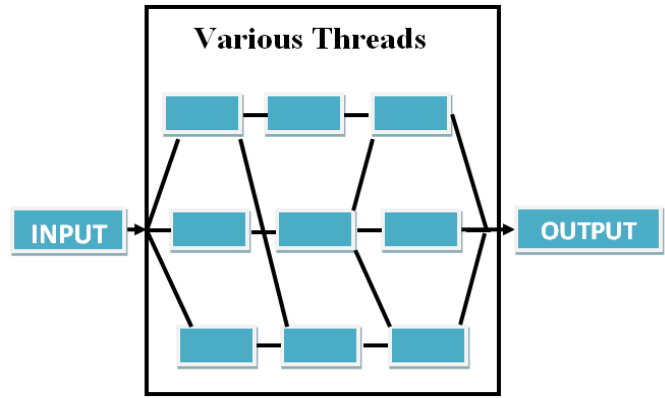

Fig. 2. Parallel Processing Architecture

II. SYSTEM ALGORITHMS AND DESCRIPTION

This section focuses on designing algorithms needed to implement parallel self sorting object.

*A. Objects' Architecture*

Each object contains two pointers and three buffers. The pointers are called prevPointer and nextPointer. The prevPointer points to the preceding or previous object and the nextPointer points to the next or the following object. The sorting algorithm forms part of the object's behaviors (or methods as used in Java). The sorting algorithm takes as parameter the sort key. This allows the object to be generalized and adaptive to different situations. With many attributes of the object, any of the attributes can be used as a sort key. The sort key is passed from one object to another object which automatically triggers the sorting behavior of the object as soon as it receives the sort key. The figure below shows the architecture of two objects in a sorting list. Sorting list refers to the set of objects to be sorted. Sorting list must contain at least two objects before sorting can take place

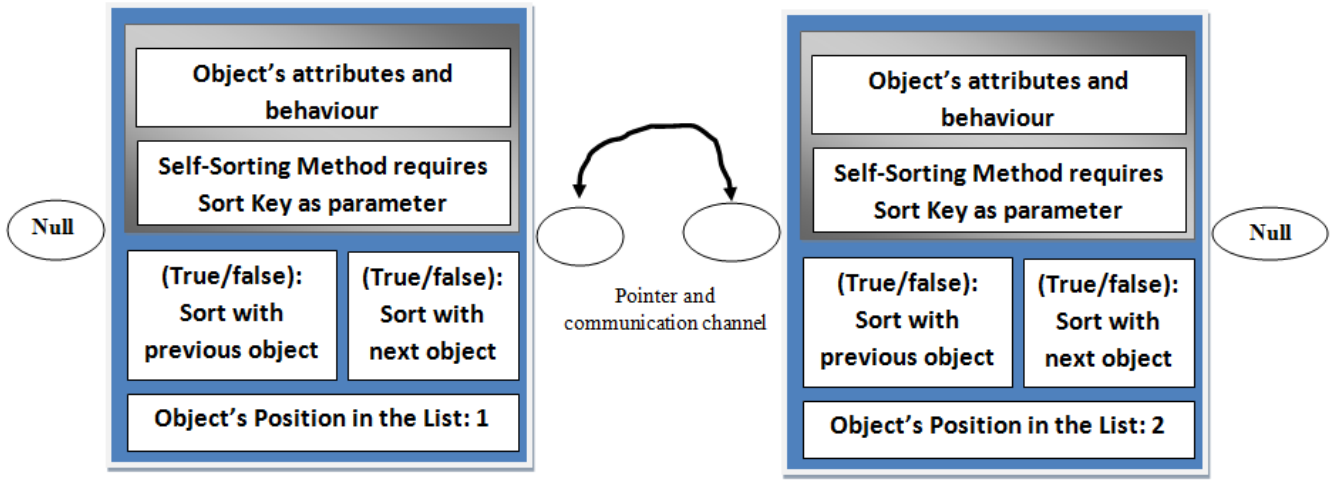

Fig. 3. Overview of Object's Architecture

The three buffers, in this work, are called prevBuffer, posBuffer and nextBuffer. Their functions in the objects are described as:

- prevBuffer: contains true when the object is sorted with the preceding or previous object based on the sort key value otherwise it contains false.
- posBuffer: contains the position of the object in the sorting list. This is updated any time the object changes position to reflect its current position. This is done by swapping the values in these buffers of the objects whose positions are changed. Initial position is assigned to an object when it is added to the sorting list.
- nextBuffer: contains true when the object is sorted with the next or the following object based on the sort key value otherwise it contains false.

The object automatically stops sorting when both prevBuffer and nextBuffer contains true. However, if a neighboring object is unsorted and wants to change position with an object which has both prevBuffer and nextBuffer true, it passes the sort key and its identity which triggers sorting, by setting one of the Buffers to false, in that object and both objects interchange positions. Interchanging position automatically triggers others to compare their sort key values with their new neighbors to check for the possibility of resorting. Eventually sorting is completed when all prevBuffers and nextBuffers are set to true.

B. *Description of System Algorithm*

The self-sorting object algorithm is said to be parallel since each object initiates a thread to get itself into the correct position. The algorithm can be considered as divide and conquer algorithm such that every two adjacent objects can be considered as being partitioned with any one being assumed as the pivot element. Combing a number of these divide and conquer objects together to get the overall list sorted describes the algorithm as Merge sort algorithm. Bubble sort concept is also said to be employed in that an object compares itself with the next object and swaps positions when needed. The object after swapping also compares itself with the next object and if required, it repeats the swapping process. Hence the parallel self-sorting algorithm implemented in this work combines concepts from different conventional sorting algorithms.

C. *Parallel Self-Sorting Design*

Every object is seen as a constituent of attributes (data fields) and behaviors (methods). These attributes and behaviors are described in terms of characters. Group of character sets together form string which can be manipulated and compared. An impulse from the surrounding environment (as part of the program when implemented in a software system) initiates sorting of the objects in the sorting list. The attribute or data field required as sort key is also provided by the impulse. The sort key is compared with the attribute list. If it exists, then sorting can be started otherwise ignore trigger. Every object starts its thread and calls its connectTo() behavior or method. The thread loops indefinitely waiting for a connection from the next or the following object in the sorting list. The Java-like algorithm for the part of the implementation of the thread is illustrated below:

```
/*this section is implemented under
run() method of each object */
try{
    {
        //list of other codes
    }
    while(true){
        conn = server.accept();
        new ConnectionHandler(conn);
    }
}catch(Exception e){//display error message}
```

Fig. 4. Overview of Connection Thread

The ConnectionHandler (conn) opens a new thread which communicates with the connected object. They exchange their sort key values and such other parameters as the value in the posBuffer later in the communication. The object also uses the connectTo() method to connect to the preceding object in the sorting list and check whether the sort key value of the preceding object is greater than its sort key value. Each object thus acts as a server and a client. If the two connected objects are not following the natural order based on their sort key values, they interchange the values in their posBuffer's and then use these values to interchange positions. The figure below demonstrates a Java-like algorithm for the general overview of the implementation:

```
if (prevPointer == null && nextPointer == null){
    //implies only one element is in the sorting list.
    //Sorting is un-necessary. Ignore trigger and stop
    ignoreTrigger();
}
if (prevPointer == null && !nextPointer == null){
    //implies that the object is the first element
    //in the sort list
    prevBuffer = true;
    nextThread.start();
}
if ( !prevPointer == null && nextPointer == null){
    //implies that the object is the last element in
    //the sort list
    nextBuffer = true;
    prevThread.start();
}
if ( !prevPointer == null && !nextPointer == null){
    //implies that the object is in the middle of
    //the sort list
    prevThread.start();
    nextThread.start();
}
if (prevBuffer == true && nextBuffer == true){
    //implies the object is sorted in relation to its
    //current position and sorting is suspended
    //stops all sorting threads for the object
    suspendSorting();
}
```

Fig. 5. General Overview of Implementation

When the object under consideration, based on the sort key value, is less than the preceding object and also greater than the next object, the value in the posBuffer received from the preceding object is given to the next object as the value from the object and also gives the value in the posBuffer received from the next object to the preceding object as the value from the object. This allows the neighboring objects of the object under consideration to swap leaving the object under consideration in its correct position. The figure below illustrates the phenomenon with an example:

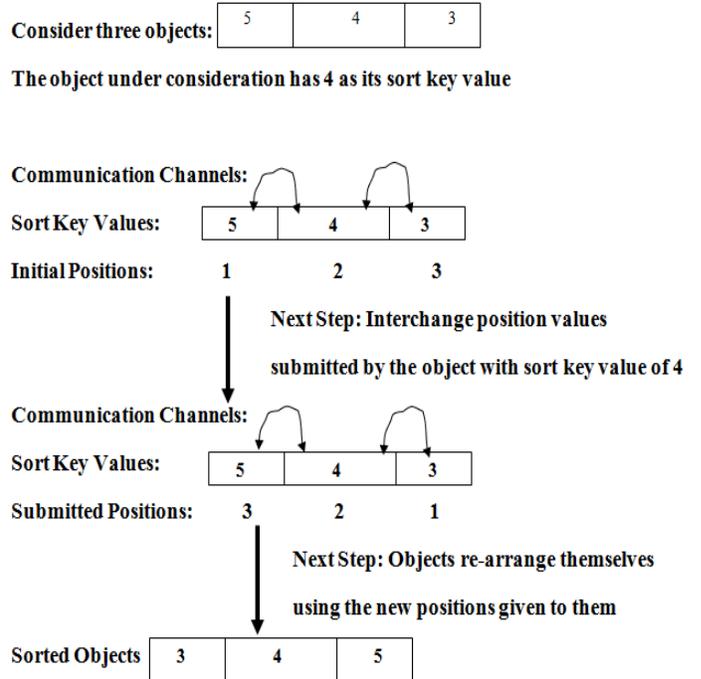

Fig. 6. Example of Swapping Non-Adjacent Objects

Swapping ensures that the new connecting neighbors update themselves to reflect the status of their neighbors in the sorting list. Take for example, a sorting list consisting of A, B, C, D, E in that order. If C and D are swapped, adjacent objects, B and E also need to update themselves. However, when B and D are swapped, non-adjacent objects, then A, C and E also need to update themselves. The Java-like algorithm that illustrates how objects manipulate sort key values is shown below:

```java
//this implementation is under a synchronised
//method which is executed when
//prevThread.start() is invoked
if (prevPointer != null &&
    compareTo(this.sortKeyValue) >
        compareTo(receivedPrevSortKeyValue)){
    //the objects follow their nature order
    //based on the sort key values
    try{wait();}catch(Exception e){}
} else if (int(this.posBuffer)-1 >= 1 &&
    compareTo(this.sortKeyValue) <
        compareTo(receivedPrevSortKeyValue)){
    swapObject(movedToPosition);
    updateConnection();
}else{
    //no sorting required
    prevBuffer = true;
}
```

Fig. 7. Preceding Object Manipulation of Sort Key Values

```java
//this implementation is under a synchronised
//method which is executed when
//nextThread.start() is invoked
if (nextPointer != null &&
    compareTo(this.sortKeyValue) <
        compareTo(receivedNextSortKeyValue)){
    //the objects follow their nature order
    //based on the sort key values
    try{wait();}catch(Exception e){}
} else if (nextPointer != null &&
    compareTo(this.sortKeyValue) >
        compareTo(receivedNextSortKeyValue)){
    swapObject(movedToPosition);
    updateConnection();
}else{
    //no sorting required
    nextBuffer = true;
}
```

Fig. 8. Next Object Manipulation of Sort Key Values

Consider an interesting scenario. Imagine that objects in a sorting list have the following sort key values in order: 6, 5, 4 and 3. The object with a sort key value 5 will initiate swapping with the object with the sort key value 6 and the object with the sort key value 4. Similarly, the object with the sort key value 4 will also initiate swapping with the object with the sort key value 5 and the object with the sort key value 3. This situation is called deadlock. Deadlock prevents swapping from taking place. swapObject(movedToPositon) method is implemented such that deadlock situations are handled successfully. The Java-like algorithm shown below illustrates part of the swapObject(movedToPosition) implementation

```java
if (compareTo(this.sortKeyValue) >
    compareTo(receivedNextSortKeyValue)  &&
        compareTo(this.sortKeyValue) <
            compareTo(receivedPrevSortKeyValue)){
    //object has to remain in its position.
    /*Save interaction parameters*/
    recNextSortKeyValue = receivedNextSortKeyValue;
    srtKeyVal = sortKeyValue;
    recPrevSortKeyValue = receivedPrevSortKeyValue;
    ignoreTrigger(); //ignore swapping
    reconnect(){
        {/*othere codes*/}
        //new information is compared with the old ones.
        //if they are the same then there is a deadlock.
        //This is solved by allowing objects to swap
        //with the preceding objects
        if (recNextSortKeyValue == receivedNextSortKeyValue &&
            srtKeyVal == sortKeyValue &&
            recPrevSortKeyValue == receivedPrevSortKeyValue){
            //swap with preceding object by
            //making the next object values very large
            inflateNextObjectParameters(receivedNextSortKeyValue);
            swapObject(movedToPosition);
        }
    };
```

```
}else{
    //move to the position using tempVariable and then swap
    if (tempVariable == null){
        //object is moved to temporary position
        tempVariable = this;
        //set the current position to null so
        //that the other object can move to occur that place
        Position(posBuffer) = null;
    }else if (Position(movedToPosition) == null){
        //the object has moved to the temporary variable
        //and the object has to occupy this positon
        Position(movedToPosition) = this;
    }
}
```

Fig. 9. Part Implementation of swapObject

## III. SYSTEM IMPLEMENTATION AND TESTING

The system was implemented and tested in various scenarios. A typical scenario in which the parallel self-sorting was used is described.

### A. Self-Sorting Implementation

The main class for implementing self-sorting object is shown in the figure below. The figure ignores data fields and private methods

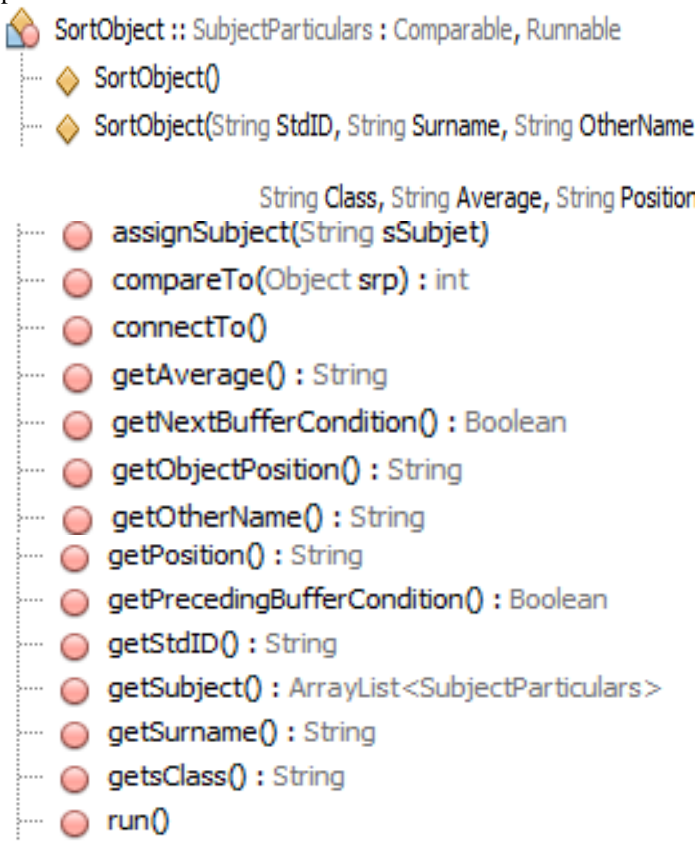

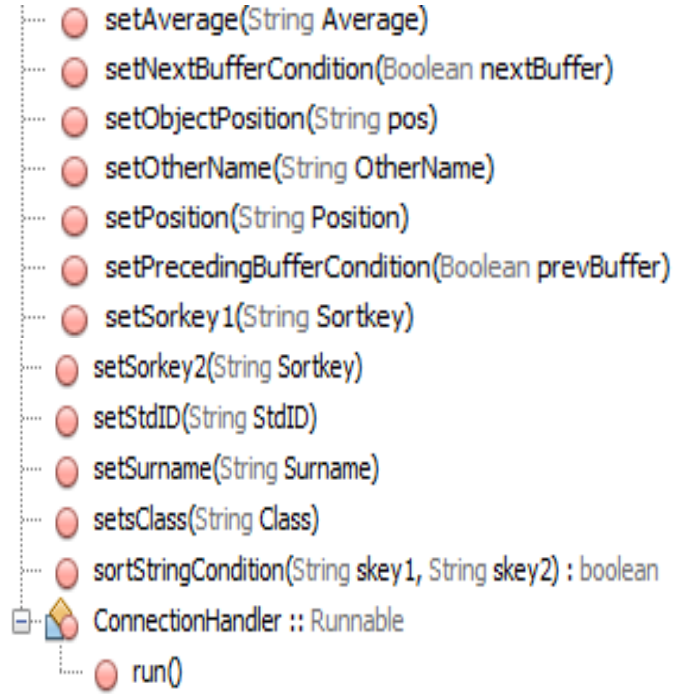

Fig. 10. Overview of Self-Sorting Object's Architecture

The object to be sorted inherits the SubjectParticulars.java class. Each object has a set of subjects and therefore the various subjects for each object is stored using arraylist, a data structure that behaves like an array but can have varying size. The figure below demonstrates the implementation of the SubjectParticulars.java class ignoring all the data fields

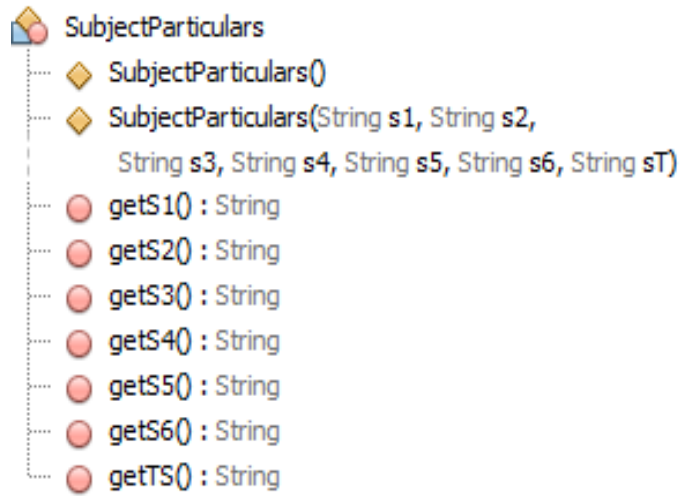

Fig. 11. Overview of SubjectParticulars.java

### B. Implementation of Testing System

The self-sorting object is implemented as part of SMSYSTEM for testing. SMSYSTEM is a system developed to support academic institutions in carrying their daily activities. SMSYSTEM represents School Management System. The system generates: terminal report, transcript, continue assessment, terminal summary report, terminal detailed report, issue students' school bill and payment statement. Self-sorting object was implemented as part of the detailed report processing. The activities in processing detailed report involve: fetching names; various subjects; class continuous assessment marks obtained through exercises, quizzes, projects and tests and exam marks from a database. All the fields of the database are string data types since a mark could contain dash (-). The system then computes the end of term or semester mark for each subject; each student's average mark; each student's position. The output involved student ID, name, subjects which are divided into core and elective, total subjects taken by each student, average mark and position of each student. The output is either formatted by names or positions. The figure below illustrates the interface needed to initiate detailed report processing.

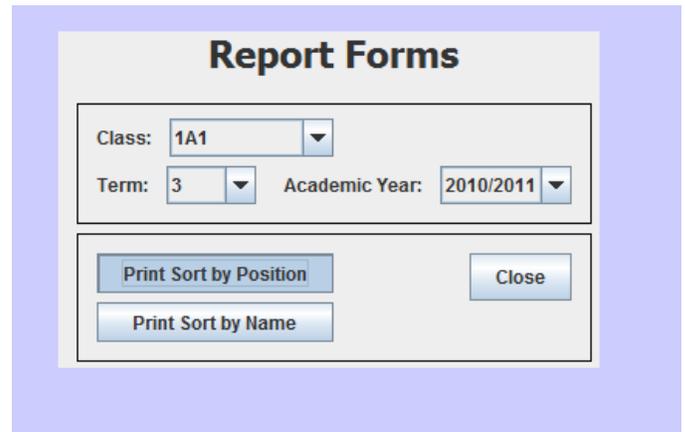

Fig. 12. Interface for Producing Detailed Report

Depending on the choice of the user, the output of the detailed report is displayed using Jasper Report. The figure below illustrates portion of the report displayed using position as a sort key.

### AGOGO STATE SNR. HIGH SCHOOL
P. O. BOX 25, ASANTE AKIM, TEL: 03220-95693

#### CLASS DETAILED REPORT

Class: 1A1   Term: 3   Academic Year: 2010/2011   No. of Students: 79   Date: 30-Nov-2011
Class Average: 51.42

| NO | STD ID | NAME | CORE SUBJECTS ||||||  ELECTIVE SUBJECTS |||| | T. SUB | AVE | POS |
|---|---|---|---|---|---|---|---|---|---|---|---|---|---|---|---|---|
| | | | MATH | ENG | ICT | SCI | PE | SS | ECO | GOV | HIS | TWI | | | | |
| 1 | 370-10 | ADJEI Francis | 58.0 | 65.0 | 56.0 | 65.0 | 72.0 | 78.0 | 81.0 | 81.0 | 72.0 | 63.0 | | 10 | 69.10 | 1ST |
| 2 | 571-10 | NTIAMOAH James | 65.0 | 51.0 | 49.0 | 72.0 | 85.0 | 77.0 | 71.0 | 71.0 | 63.0 | 75.0 | | 10 | 67.90 | 2ND |
| 3 | 162-10 | ISSAH Fati | 41.0 | 62.0 | 61.0 | 62.0 | 83.0 | 85.0 | 66.0 | 57.0 | 67.0 | 82.0 | | 10 | 66.60 | 3RD |
| 4 | 355-10 | BOSOMPIM Sylvia | 76.0 | 54.0 | 46.0 | 66.0 | 74.0 | 74.0 | 50.0 | 70.0 | 66.0 | 76.0 | | 10 | 65.20 | 4TH |
| 5 | 527-10 | KONADU Gladys | 61.0 | 56.0 | 49.0 | 65.0 | 82.0 | 75.0 | 61.0 | 70.0 | 63.0 | 68.0 | | 10 | 65.00 | 5TH |
| 6 | 497-10 | AMANING Samuel | 55.0 | 59.0 | 40.0 | 70.0 | 74.0 | 67.0 | 85.0 | 62.0 | 67.0 | 62.0 | | 10 | 64.10 | 6TH |
| 7 | 064-10 | DARKO Jeniffer | 54.0 | 63.0 | 45.0 | 63.0 | 71.0 | 82.0 | 72.0 | 52.0 | 67.0 | 67.0 | | 10 | 63.60 | 7TH |
| 8 | 010-10 | ASARE KOFI Emmanuel | 70.0 | 52.0 | 53.0 | 58.0 | 75.0 | 73.0 | 61.0 | 63.0 | 59.0 | 70.0 | | 10 | 63.40 | 8TH |
| 9 | 487-10 | OKYERE Peter | 48.0 | 60.0 | 48.0 | 59.0 | 84.0 | 75.0 | 70.0 | 65.0 | 55.0 | 66.0 | | 10 | 63.00 | 9TH |
| 10 | 248-10 | OSEI Jephter Akoto | 56.0 | 55.0 | 39.0 | 64.0 | 70.0 | 76.0 | 81.0 | 63.0 | 48.0 | 71.0 | | 10 | 62.30 | 10TH |
| 11 | 368-10 | ANAKWA Amos | 51.0 | 60.0 | 45.0 | 58.0 | 70.0 | 82.0 | 70.0 | 70.0 | 59.0 | 57.0 | | 10 | 62.20 | 11TH |
| 12 | 048-10 | ASAMOAH Dosi Augustine | 44.0 | 57.0 | 53.0 | 51.0 | 67.0 | 74.0 | 74.0 | 60.0 | 68.0 | 59.0 | | 10 | 60.70 | 12TH |
| 13 | 363-10 | MUSTAPHA Faiza | 52.0 | 71.0 | 42.0 | 63.0 | 81.0 | 69.0 | 61.0 | 50.0 | 57.0 | 57.0 | | 10 | 60.30 | 13TH |
| 14 | 087-10 | TETTEH Charity | 41.0 | 57.0 | 45.0 | 56.0 | 74.0 | 74.0 | 71.0 | 56.0 | 61.0 | 61.0 | | 10 | 59.60 | 14TH |
| 15 | 256-10 | OPOKU Michael | 56.0 | 52.0 | 55.0 | 62.0 | 66.0 | 64.0 | 53.0 | 56.0 | 62.0 | 70.0 | | 10 | 59.60 | 14TH |
| 16 | 031-10 | AFRIYIE Elvis | 39.0 | 48.0 | 44.0 | 69.0 | 57.0 | 68.0 | 59.0 | 67.0 | 73.0 | 69.0 | | 10 | 59.30 | 16TH |
| 17 | 060-10 | BOADI NKETSIAH Harriet | 52.0 | 64.0 | 42.0 | 57.0 | 69.0 | 70.0 | 62.0 | 60.0 | 46.0 | 69.0 | | 10 | 59.10 | 17TH |
| 18 | 202-10 | ATAKORA Priscilla | 49.0 | 62.0 | 45.0 | 55.0 | 69.0 | 68.0 | 67.0 | 45.0 | 59.0 | 71.0 | | 10 | 59.00 | 18TH |

Class Teacher's Signature:................................................................

Page 1 of 4

Fig. 13. Position Sort Key Sample of Detailed Report

*C. Analysis of the Implemented System*

At the baseline, the efficiency of the system when implemented as software depends on the capacity of the memory and the processing power of the system. However, if an object with one attribute is implemented and compared with the conventional sorting algorithms whose concepts were employed, the parallel self-sorting system proposed in this research, in the best case, works at the speed of $O(n \log n)$. Concerning the worst case, the algorithm has $O(n^2 \log n)$ complexity.

The system is very effective in sorting objects when it was tested in software implementation. Regardless of the number of attributes of different objects, the system only requires that all the objects to be sorted have a common sort key which is among the attributes of each object.

## IV. CONCLUSION

In this paper, the researcher outlined the various techniques required to implement a parallel sorting system whose concept is based on the combination of techniques from the conventional sorting algorithms. The effectiveness and efficiency of the proposed system were verified through the use of different objects with variable attributes and behaviors. Testing results show that the system is effective regardless of different number of attributes. Its efficiency, however, depends on the capacity of the memory and processing power when implemented in software environment.

The system serves as a basis for designing and implementing self sorting applications in sensitive situations such as arrangement of linear sequence communicating nodes to provide optimal performance and arrangement of autonomous systems, environment sensors or intelligent agents in a particular order to perform designated functions.